\documentclass[twocolumn,aps,prb,groupedaddress,showpacs]{revtex4}
\usepackage{mathrsfs}
\usepackage{bm}
\usepackage{multirow}
\usepackage{amsmath,amsfonts,amssymb}
\usepackage{array,booktabs}
\usepackage{graphicx,times,graphics,color,epsfig}
\usepackage{enumerate}

\begin{document}
\title{Statistical properties of electrochemical capacitance in disordered mesoscopic capacitors}
\author{Fuming Xu}
\affiliation{Department of Physics and the Center of Theoretical and Computational Physics, The University of
Hong Kong, Hong Kong, China.}
\author{Jian Wang}
\email{jianwang@hku.hk}
\affiliation{Department of Physics and the Center of Theoretical and Computational Physics, The University of
Hong Kong, Hong Kong, China.}

\begin{abstract}
We numerically investigate the statistical properties of electrochemical capacitance in disordered two-dimensional mesoscopic capacitors. Based on the tight-binding Hamiltonian, the Green's function formalism is adopted to study the average electrochemical capacitance, its fluctuation as well as the distribution of capacitance of the disordered mesoscopic capacitors for three different ensembles: orthogonal (symmetry index $\beta$=1), unitary ($\beta$=2), and symplectic ($\beta$=4). It is found that the electrochemical capacitance in the disordered systems exhibits universal behavior. In the case of single conducting channel, the electrochemical capacitance follows a symmetric Gaussian distribution at weak disorders as expected from the random matrix theory. In the strongly disordered regime, the distribution is found to be a sharply one-sided form with a nearly-constant tail in the large capacitance region. This behavior is due to the existence of the necklace states in disordered systems, which is characterized by the multi-resonance that gives rise to a large density of states. In addition, it is found that the necklace state also enhances the fluctuation of electrochemical capacitance in the case of single conducting channel. When the number of conducting channels increases, the influence of necklace states becomes less important. For large number of conducting channels, the electrochemical capacitance fluctuation develops a plateau region in the strongly disordered regime. The plateau value is identified as universal electrochemical capacitance fluctuation, which is independent of system parameters such as disorder strength, Fermi energy, geometric capacitance, and system size. Importantly, the universal electrochemical capacitance fluctuation is the same for all three ensembles, suggesting a super-universal behavior.
\end{abstract}

\pacs{72.10.-d,
73.21.La,
73.23.-b,
73.63.-b
}

\maketitle

\section{introduction}

As the basic building block of nanoscale electronic circuit, nanoscale capacitor has been the focus of intense study since its fundamental significance. Due to the small sample sizes, the finite density of states (DOS) in nanostrutures can give rise to a non-negligible quantum correction to their physical properties. For instance, a mesoscopic capacitor shows different dynamic response to an ac external bias from its classical counterpart. At low frequencies, the transport properties of a mesoscopic capacitor is characterized by the electrochemical capacitance $C_\mu$, charge relaxation resistance $R_q$, and quantum inductance.\cite{buttiker93,jian07PRB}
Numerous research efforts have been devoted in the study  of mesoscopic capacitor in the past
decade.\cite{macucci95JAP,buttiker96PRL,buttiker99JKPS,jian98PRL,carlo05PRB,buttiker06PRL,buttiker07Nano,
buttiker08PRL,buttiker08PRB,mora10NPHYS,expt01,expt06,expt06NP,expt10,expt12}

It is known that for a bulk metal, the huge density of states results in a very small screening length while for a mesoscopic conductor, the finite density of states gives rise to a finite screening length. As a result, the electrochemical potential controlled experimentally is not equal to the electrostatic potential that defines the classical capacitance. This leads to a quantum correction to the classical capacitance that has been confirmed experimentally.\cite{quote1} This electrochemical capacitance $C_\mu$ can be treated in a dynamic way using scattering matrix theory\cite{buttiker93} or non-equilibrium Green's function.\cite{wei1} At low frequencies, $C_\mu$ is determined by the serial combination of the geometrical capacitance and the quantum capacitance which is induced by the finite DOS. It was shown that $C_\mu$ oscillates as a function of the Fermi energy.\cite{buttiker07Nano} In addition, the dissipative part of a mesoscopic coherent capacitor is related to the charge relaxation resistance which was predicted to be a universal quantity equal to half the resistance quantum $R_q=h/2e^2$ for single conducting channel.\cite{buttiker93} The universal value of charge relaxation resistance is independent of Fermi energy even in the presence of strong electron electron interaction.\cite{buttiker06PRL,expt06} Recently, an experimental study on charge relaxation resistance was carried out on a mesoscopic capacitor made of a two dimensional electron gas (2DEG) quantum dot with the second metallic plate on top of it.\cite{expt06} This setup is equivalent to a RC circuit consisting of classical capacitance, quantum capacitance and the charge relaxation resistance. By measuring the frequency dependent conductance of this RC circuit in GHz regime, the charge relaxation resistance was obtained and found to be half of the resistance quantum in the single conducting mode. \cite{expt06} At higher frequency, quantum inductance becomes relevant and the effective capacitance of the mesoscopic capacitor can be negative,\cite{jian07PRB} which has been seen experimentally in gated carbon nanotubes.\cite{carlo05PRB}

In mesoscopic systems, the sample-to-sample fluctuation of physical quantities such as conductance exhibits universal behavior in the diffusive regime. This universal conductance fluctuation (UCF) is one of the hallmarks of mesoscopic physics. It was known that value of UCF depends only on the symmetry and dimensionality of the systems. In the language of random matrix theory, the symmetry is characterized by symmetry index $\beta$ where $\beta$=1,2,4 corresponds to orthogonal, unitary, and symplectic ensembles. Not only the conductance fluctuation has universal values in the diffusive regime, the conductance distribution also exhibits universal features in the localized regime.\cite{distr,qiao10PRB} Ac transport properties of mesoscopic systems also show sample to sample fluctuations in disordered mesoscopic systems. Study on the ac admittance of a chaotic quantum dot using random matrix theory (RMT) revealed that the admittance is governed by two time scales.\cite{brouwer} The emittance fluctuation of mesoscopic conductors were investigated including 1D, 2D, and quantum dot systems. It was found that due to the existence of the necklace states, the average emittance is negative from ballistic regime all the way to the localized regime showing inductive-like behavior.\cite{ren0} Using RMT,\cite{beenakker97RMP} the distribution functions of the electrochemical capacitance for a chaotic quantum dot system with single conducting channel were derived for three symmetry classes $\beta$=1,2,4.\cite{buttiker96PRL} To best of our knowledge, there is no theoretical study of electrochemical capacitance in disordered systems when the number of conducting channels is large. Therefore, it is interesting to conduct such an investigation. In particular, we are interested in the distribution function of $C_\mu$ since there are theoretical results\cite{buttiker96PRL} to compare with. Also, inspired by the universal conductance fluctuation\cite{UCF} in diffusive regime of disordered systems, we are interested in exploring the fluctuation behavior of electrochemical capacitance in different symmetry classes.

In this paper, we perform an extensive numerical investigation on the statistical properties of a mesoscopic capacitor shown schematically in Fig.\ref{fig01}. In the numerical calculation, the tight-binding model is used with Anderson-type disorder to realize different random configurations. We study electrochemical capacitance fluctuation and its distribution function for three ensembles with $\beta$=1,2,4. For the case of single conducting channel, it is found that the capacitance distribution shows symmetric Gaussian shape in the weak disorder regime and changes to a one-sided form centered in large capacitance region as the disorder is increased. At strong disorders, the distribution spreads out in the whole range of capacitance and eventually peaks in small capacitance area with a long tail that corresponds to the large density of states. This long tail can not be accounted for within random-matrix theory.\cite{buttiker96PRL} It is due to the existence of necklace states in disordered systems which are characterized by multi-resonance with large density of states.\cite{azbelresonance} As a result, the fluctuation of electrochemical capacitance is greatly enhanced at single conducting channel case, which is much larger than the theoretical value predicted by RMT.\cite{buttiker96PRL} As the number of conducting channels $N$ increases, the necklace states become less prominent and the fluctuation of electrochemical capacitance develops a plateau structure at large disorder strength. The electrochemical capacitance fluctuation at large $N$ is found to be an universal value, which is independent of the disorder strength, Fermi energy, geometric capacitance and system size. Remarkably, this universal capacitance fluctuation is the same for all three ensembles. This super-universal behavior is very different from UCF, whose value depends on the symmetry index. The distribution function of electrochemical capacitance also exhibits universal features. We find that at large $N$ the capacitance distribution is a universal function that depends only on the average capacitance and does not depend on the symmetry of the system.

The paper is organized as follows. In Sec.\ref{sec2}, the theoretical formalism and its numerical implementation are presented. In Sec.\ref{sec3}, we show the numerical results on the distribution and fluctuation of electrochemical capacitance for various disorder strengths, Fermi energies, geometric capacitances and three different ensembles, accompanied with discussion and analysis. Finally, the conclusion is drawn in Sec.\ref{sec4}.

\section{theoretical formalism}\label{sec2}

\begin{figure}[tbp]
\centering
\includegraphics[width=0.8\columnwidth]{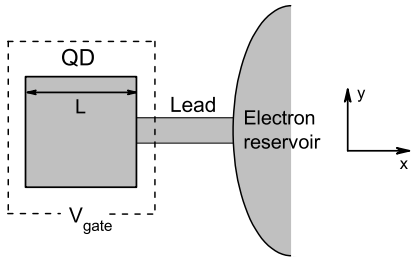}
\caption{Sketch of a parallel plate mesoscopic capacitor constructed
on 2DEG. Square region of width $L$ is a mesoscopic plate. A single
narrow lead whose width is $L/4$ connects this plate to an electron
reservoir. Dash-dot line stands for a large classical plate which
acts as another electrode that the mesoscopic plate is capacitively
coupled with through the applied gate voltage $V_{gate}$. }
\label{fig01}
\end{figure}

The parallel plate capacitor considered in this work is shown schematically in Fig.\ref{fig01} which contains one mesoscopic plate of size $1600nm \times 1600 nm$ referred as quantum dot (QD). A semi-infinite lead of width $400 nm$ connects this plate to the electron reservoir. The other plate of the capacitor is a large metallic plate on the top of the mesoscopic plate. It has a large DOS so that the quantum correction due to DOS can be neglected.
The Fermi energy of the mesoscopic plate can be tuned by a gate voltage $V_{gate}$. According to Ref.[\onlinecite{buttiker93}], the electrochemical capacitance of this system is defined as
\begin{equation}
\frac{1}{C_\mu}= \frac{1}{C_e}+ (\frac{e^2}{2 \pi i} Tr[S^{\dag}
\frac{dS}{dE}])^{-1}
\end{equation}
where $C_e$ is the classical geometrical capacitance and $\mathcal {N}$= $\frac{1}{2 \pi i} {S^{\dag} \frac{dS}{dE}}$ is the Wigner-Smith time-delay matrix\cite{wignersmith} with $S$ the scattering matrix. The second term of this equation is also called quantum capacitance $C_q$, which is determined by the density of states (DOS) of the system at low frequency\cite{buttiker93} $C_q= \frac{e^2}{2 \pi i} Tr[S^{\dag} \frac{dS}{dE}]\equiv e^2 D$. The electrochemical capacitance $C_\mu$ is simply the serial combination of geometrical capacitance $C_e$ and quantum capacitance $C_q$, which reads
\begin{equation}
\frac{1}{C_\mu}= \frac{1}{C_e}+ \frac{1}{e^2 D} \label{eq01}
\end{equation}

To study the statistical properties of electrochemical capacitance in the presence of disorders, we will consider three ensembles: (1). orthogonal ensemble with $\beta=1$ in the language of random matrix theory\cite{beenakker97RMP} where both the magnetic field and the spin orbit interaction are absent. (2). unitary ensemble with $\beta=2$ where the magnetic field is nonzero while the spin orbit interaction is zero. (3). symplectic ensemble with $\beta=4$ where the magnetic field is zero and the spin orbit interaction is nonzero. We use the following general tight-binding Hamiltonian for different symmetry classes\cite{qiao10PRB}
\begin{equation}
\begin{array}{cll}
H=& \sum_{nm\sigma}(\varepsilon_{nm} +
v_{nm})c^\dag_{nm\sigma}
c_{nm\sigma} \\
& - t\sum_{nm\sigma}[ c^\dag_{n+1,m\sigma} c_{nm\sigma}
e^{-im\phi} + c^\dag_{n,m-1,\sigma} c_{nm\sigma} + h.c.] \\
& - t_{SO} \sum_{nm\sigma \sigma^{'}}[c^\dag_{n,m+1,\sigma}
(i\sigma_x)_{\sigma \sigma^{'}}c_{nm\sigma} \\
& - c^\dag_{n+1,m\sigma} (i\sigma_y)_{\sigma
\sigma^{'}}c_{nm\sigma^{'}} e^{-im\phi} + h.c.]
\end{array}\label{eq02}
\end{equation}
where $c^\dag_{nm\sigma}$($c_{nm\sigma}$) is the electron creation (annihilation) operator on the lattice site $(n,m)$. Here $\varepsilon_{nm}$ represents the on-site energy with magnitude $4t$ in the 2D system and $t$ is the nearest-neighbor hopping energy; $v_{nm}$ is the Anderson-type disorder potential which has uniform distribution in the interval [-W/2,W/2] with $W$ the disorder strength; $\sigma_{x/y}$ are the Pauli matrices and $t_{SO}$ is the spin-orbit coupling strength. The magnetic flux $\phi$ is due to the magnetic field perpendicular to the mesoscopic plate. Different symmetry classes are realized by turning on or off the magnetic field as well as the spin-orbit interaction in the Hamiltonian. For numerical calculation, we discretize the mesoscopic plate using a $80\times80$ square grid with lattice spacing $a=20nm$, while the semi-infinite lead has a width of 20 sites along y-direction (see Fig.1). In this system we set the hopping energy $t$ as the energy unit with $t=\hbar^2/{2m^* a^2}= 1.42 meV$, where the effective mass $m^*$ is typically that of GaAs, $m^*=0.067 m_e$. The geometric capacitance is chosen to be $C_e=17.7fF$, which is reasonable for a parallel plate capacitor with size $1600\times 1600 {nm}^2$.

In terms of the Green's function, DOS can be expressed as
\begin{equation}
{\rm D} = \frac{1}{2\pi} {\rm Tr}[G^r\Gamma G^a]
\end{equation}
where $G^r=[E-H-\Sigma^r]^{-1}$ is the retarded Green's function, $G^a=(G^r)^\dagger$, $\Sigma^r$ is the self-energy of the lead, and $\Gamma$ is the line width function describing coupling of the lead to the quantum dot with $\Gamma=i[\Sigma^{r} -\Sigma^{a}]$. In the following, we will study the dimensionless electrochemical capacitance $\alpha$ which is the ratio between the electrochemical capacitance and the electrostatic (geometrical) capacitance, $\alpha=C_\mu/C_e$. Obviously, we have $0<\alpha<1$ since $D$ is positive definite. For systems with very large DOS we have $\alpha \sim 1$ while for systems with very small DOS we have $\alpha \sim e^2 D/C_e$.

In the presence of disorder, the average electrochemical capacitance $\alpha$ and its fluctuation are defined as
\begin{align}
&<\alpha>=<C_\mu/C_e> \notag \\
&rms(\alpha)=\sqrt{<(\alpha-<\alpha>)^2>}\notag
\end{align}
where $<\cdots>$ denotes the averaging over different random configurations with the same disorder strength $W$.

\section{numerical results and discussion}\label{sec3}

In this section, numerical results on the statistical properties of dimensionless capacitance $\alpha$ are presented. We will first discuss the distribution function of $\alpha$ and make comparison with existing theory and then study the average capacitance and its fluctuation.

\subsection{The distribution of electrochemical capacitance}\label{sec31}

\begin{figure*}[tbp]
\centering
\includegraphics[width=17cm]{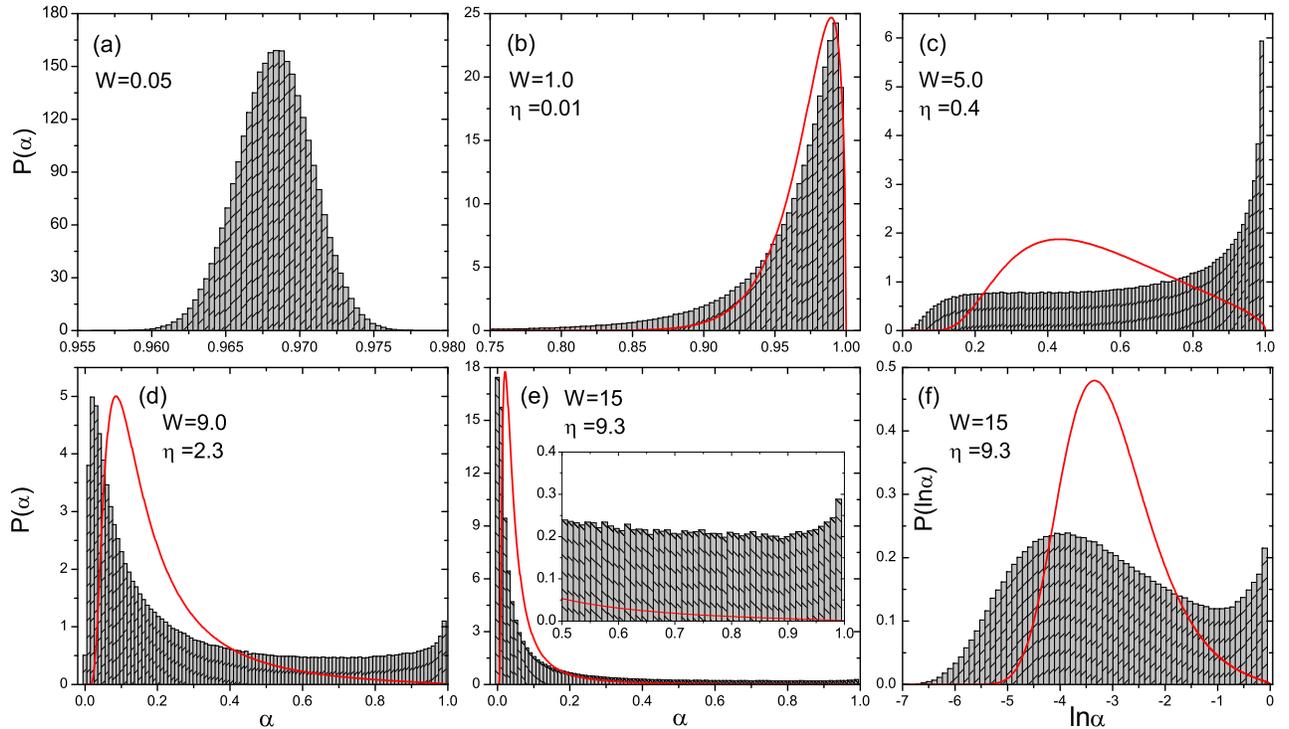}
\caption{Distribution of the dimensionless capacitance $\alpha=C_\mu/ C_e$ for orthogonal ensemble with single conducting channel ($N$=1) in the lead. We choose $E$=0.065 in the calculation. The gray histograms in all the panels are from numerical calculation, corresponding to different disorder strengths, while the red lines are theoretically predictions\cite{buttiker96PRL} at various values of parameter $\eta$. The theoretical results are shown for qualitative comparison.}\label{fig02}
\end{figure*}

We first study the statistical properties of electrochemical capacitance when only one conducting channel is available, since it has been experimentally realized\cite{expt06} and theoretically investigated from the random-matrix theory.\cite{buttiker96PRL} From the random matrix theory, the distribution of $\alpha=C_\mu/C_e$ for a general symmetry class with symmetry index $\beta$ in the case of single conducting channel reads\cite{buttiker96PRL}
\begin{equation}
P_{\beta,\eta}(\alpha)=\frac{(\beta/2)^{\beta/2}}
{\Gamma(\beta/2)}\frac{(1-\alpha)^{\beta/2}}{\eta^{(\beta+2)/2}
\alpha^{(\beta+4)/2}} e^{-\beta(1-\alpha)/{2\eta
\alpha}}\label{eq03}
\end{equation}
where $\eta=C_e\Delta/e^2$ is a dimensionless parameter with $\Delta$ the mean level spacing of the disordered sample. From its definition,$\eta$ is inversely proportional to the level density and hence goes to zero for macroscopic sample. In addition $\eta$ contains geometric information of the system through geometric capacitance Ce. In Sec.\ref{sec32}, we find that $\eta$ plays a similar role as disorder strength in our calculation.

We start with the orthogonal ensemble. For one incoming conducting channel $N=1$, we fix the Fermi energy at $E=0.065$. The numerically calculated distributions of electrochemical capacitance $\alpha$ at different disorder strengths are shown in Fig.\ref{fig02} where the histogram is obtained from 1,000,000 configurations at each disorder strength. The theoretical prediction from Eq.(\ref{eq03}) at different $\eta$ are shown as red lines where we have selected $\eta$ so that the resulting distribution resembles the numerical result. We see from Fig.\ref{fig02}a that, at very weak disorder $W=0.05$ $P(\alpha)$ is approximately a Gaussian distribution with the peak located at $\alpha=0.967$, indicating that the system is in the ballistic regime. As increasing the disorder strength, the distribution shifts towards small capacitance. At $W=1$ we see the occurrence of small capacitances due to the quantum correction of DOS and $P(\alpha)$ changes gradually from Gaussian to one-sided distribution shown in panel b. A theoretical result from Eq.(\ref{eq03}) with $\eta=0.01$ shows a similar behavior. For a larger disorder strength $W=5$ (panel c), $\alpha$ distributes almost uniformly in the whole range except that there is a counter-intuitive narrow peak for $\alpha$ close to one. We found that Eq.(\ref{eq03}) is not able to give such a peak. This peak is suppressed when disorder increases further to $W=9$, and $\alpha$ starts to peak in the region of small $\alpha$, as shown in panel d. It is found that $P(\alpha)$ remains approximately constant value in the range $\alpha=(0.2,0.9) $ with a magnitude of one fifth of the peak value. On the other hand, the theoretical curve decreases rapidly at large $\alpha$. Upon further increasing the disorder ($W=15$ in panel e), the system is localized.\cite{localization} At this disorder strength $P(\alpha)$ is sharply peaked at small value of $\alpha$ with a long tail stretching to $\alpha=1$. From the inset of Fig.\ref{fig02}e we see that this tail has approximately a constant distribution in range $\alpha=[0.5,1]$ which clearly deviates from Eq.\ref{eq03}. In Fig.\ref{fig02}f we draw the distribution of the logarithmic of $\alpha$ at this disorder, where an abnormal peak near $\ln(\alpha) \simeq 0$ appears\cite{fmx08thesis}.

From Fig.2, we notice two interesting phenomena that disagrees with the theoretical prediction. One is the counter-intuitive narrow peak around $\alpha$=1 at $W=5$, which means that some disorder configurations have very large DOS making quantum capacitance very small. The other is the unexpected long tail in the strong localized regime with a nearly constant distribution probability at large $\alpha$, where $P(\alpha)$ is theoretically expected to decay. We attribute these counter-intuitive results to the existence of Azbel resonance states or necklace states,\cite{azbelresonance} which exist in strongly disordered systems as a result of multi-resonance processes. These states are rare events and form a small portion of the whole ensemble.\cite{pendry87JPC} Although rare, these states can have significant contribution to the ac transport properties of disordered systems. Due to the nature of resonant process, the necklace states (or precursor of necklace state in the diffusive regime) give rise to large DOS. We note that the existence of necklace states was confirmed experimentally in a quasi-1D optical system.\cite{bertolottiPRL05}  In Ref.[\onlinecite{xu11PRB}], it was numerically found that the necklace states widely exist in disordered 2D and QD systems for all symmetry classes $\beta$=1,2,4 and leads to the universal power-law behavior of the distribution of Wigner delay time $\tau$ at large $\tau$: $P(\tau)\sim \tau^{-2}$. Here in the distribution of electrochemical capacitance for a disordered mesoscopic capacitor, we find again the evidence of necklace states. Because of the huge DOS from the necklace states the distribution of $\alpha = C_\mu/C_e$ in Fig.\ref{fig02}c shows an unexpected large sharp peak in large $\alpha$ region and exhibits a long tail distributed in a wide range of $\alpha$ with nearly constant probability (for $W \geq 9$). It is very different from the theoretical predictions obtained from random-matrix theory\cite{buttiker96PRL} where the necklace states are not considered. The peak in Fig.\ref{fig02}f near $\ln \alpha=0$ at $W=15$ is also originated from necklace states. We wish to point out that necklace states manifest only for transport properties that are sensitive to DOS. For instance, emittance and electrochemical capacitance are significantly affected by necklace states. For conductance, however, necklace states have no effect.\cite{ren}

\begin{figure}[tbhp]
\includegraphics[width=\columnwidth]{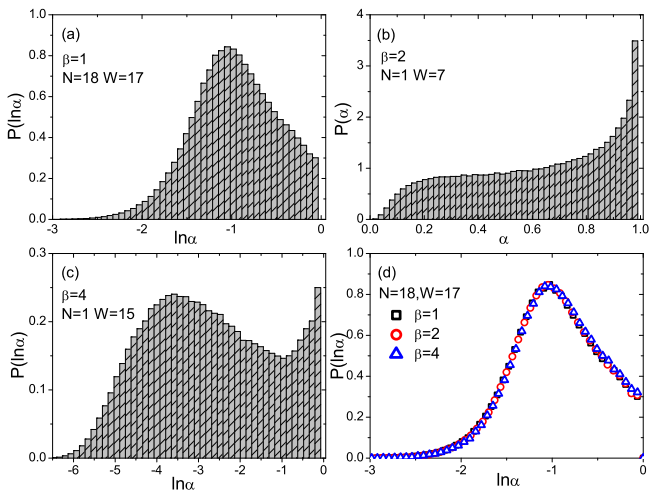}
\caption{Panel (a) plots the distribution of ${\rm ln\alpha}$ at orthogonal symmetry with $N=18$ and $W=17$. Panel (b) shows the distribution function of a system with magnetic field ($\beta=2$) at single channel and disorder strength $W=7$. Panel (c) depicts $P({\rm ln\alpha})$ at symplectic case ($\beta=4$) for $N=1$ and $W=15$. Panel (d) shows the distribution function at $N=18$ and $W=17$ for all three symmetry classes $\beta=1,2,4$. Panel (b) is almost the same as Fig.\ref{fig02}c and Panel (c) is very similar to Fig.\ref{fig02}f.}\label{fig03}
\end{figure}

Since the necklace states are due to the multi-resonance process, their influence on the distribution of electrochemical capacitance is notable at $N=1$. When there are $N$ propagating channels in the system, the longitudinal velocity of different channels are different for a given energy. As a result, the effect of necklace states is suppressed because only one channel can be on or close to resonance and has large DOS while the DOS of other channels are small. Hence, comparing with the single channel situation, the number of necklace states is expected to
decrease by a factor of $N$. Therefore we expect that the influence of necklace states on the distribution of capacitance $\alpha$ becomes much weaker for large $N$. To support this argument, we provide numerical evidence on the effect of necklace states in Fig.\ref{fig03}. In Fig.\ref{fig03}a we calculate the distribution of ${\rm ln\alpha}$ for a larger Fermi energy with conducting channel $N=18$ at disorder strength $W=17$. Comparing with Fig.\ref{fig02}f which corresponds to $N=1$ and $W=15$, it is found that the peak at large DOS near ${\rm ln\alpha=0}$ disappears. The fact that ${\rm P(ln\alpha)}$ doesn't decay to zero at this point shows that there are still necklace states at large number of conducting channels.

So far, we have discussed the distribution of $\alpha = C_\mu/C_e$ for the orthogonal symmetry with single and multiple conducting channels. Our numerical calculations on symmetry classes $\beta=2,4$ give very similar results, i.e., the distributions of electrochemical capacitance in unitary and symplectic ensembles follow similar trend as in orthogonal ensemble shown in Fig.\ref{fig02}. The necklace states are important for the case of single conducting channel ($N=1$) and are less important when $N$ becomes large. For instance, in Fig.\ref{fig03}b, the distribution function $P(\alpha)$ is generated from large number of disorder samples in the presence of magnetic field $B=0.01$ (unitary ensemble). With single channel in the lead and disorder strength $W=7$, the sharp peak in $P(\alpha)$ is very similar to that shown in Fig.\ref{fig02}c for symmetry class $\beta=1$, highlighting the importance of necklace states for $N=1$. Similar behavior is found for symplectic ensemble with single transmission channel. In Fig.\ref{fig03}c, we show the distribution of logarithmal $\alpha$ for symplectic symmetry class with $t_{SO}=0.2$. We see that $P(\ln\alpha)$ at $N=1$ also exhibits necklace states similar to Fig.\ref{fig02}f. We have also calculated distribution function $P(\ln\alpha)$ for $\beta=2,4$ when the number of conducting channel is large ($N$=18) while fixing disorder strength at $W=17$. Remarkably $P(\ln\alpha)$ for three ensembles are on top of each other showing super-universal behavior (see Fig.\ref{fig03}d). We note that for conductance fluctuation in the diffusive regime, the universal conductance fluctuation (UCF) which is obtained as the second moment of its distribution function depends only on the symmetry of the system as well as the dimensionality.\cite{qiao10PRB} Our results on capacitance distribution function suggest that the distribution function and hence the capacitance fluctuation at $W=17$ is independent of symmetry index $\beta$ at large $N$. In the next subsection, we will demonstrate that the electrochemical capacitance distribution function at large $N$ depends only on the average electrochemical capacitance and does not depend on the symmetry index.

We briefly summarize the results on the distribution of dimensionless capacitance $\alpha$. In the case of a single transmission channel or a few conducting channels, the distribution of capacitance gradually changes from Gaussian to one-sided normal distribution peaked at large $\alpha$ for weak disorders. For medium disorders, the distribution $P(\alpha)$ spreads out in the whole interval $\alpha \in (0, 1]$ with a sharp peak around $\alpha=1$. At strong disorders, the distribution function becomes one-sided again but sharply peaked at small $\alpha$, with long tail of nearly constant magnitude in large $\alpha$. Both the peak and tail in the capacitance distributions arise from the existence of necklace states in disordered systems, which give rise to large DOS. When the number of conducting channels is large, the peak and tail due to necklace states are suppressed. The distribution functions for three different ensembles collapse into a single curve at large number of conducting channels and strong disorder strengths.

\subsection{The fluctuation of electrochemical capacitance}\label{sec32}

It is well known that the conductance fluctuation in the diffusive mesoscopic system assumes an universal value\cite{UCF} that is independent of Fermi energy, disorder strength, and sample size. Similar universal behavior for the spin-Hall conductance fluctuation\cite{USCF} was also reported in disordered mesoscopic systems. It would be interesting to see whether the fluctuation of electrochemical capacitance exhibits universal features in the presence of disorders.

\begin{figure}[tbp]
\includegraphics[width=\columnwidth]{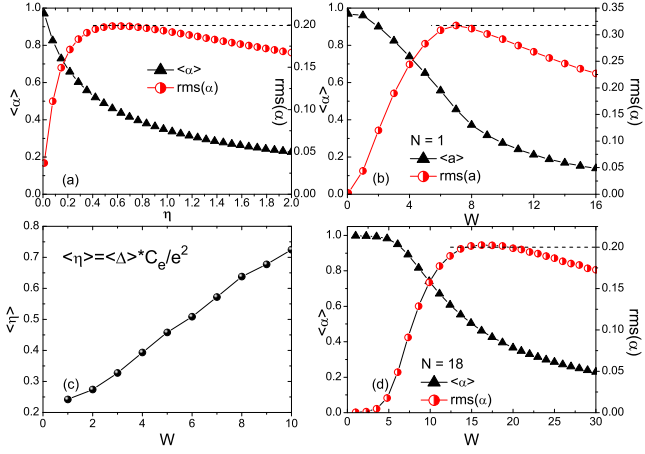}
\caption{Panel (a) plots the average capacitance and its fluctuation vs $\eta$, obtained from the theoretical prediction Eq.(\ref{eq03}). Panel (b) is the numerical result calculated at Fermi energy $E=0.065$ with channel number $N=1$ in the system. Panel (c) shows the calculated $\eta$ from mean level spacing as a function of $W$. Panel (d) corresponds to average $\alpha$ and its fluctuation versus $W$ in the case of $N=18$. All results are obtained from the orthogonal ensemble.}\label{fig04}
\end{figure}

In Fig.\ref{fig04}, we plot the average capacitance and its fluctuation vs disorder strengths for $\beta=1$ when the number of incoming channel $N=1$. For single conducting channel situation, our numerical result can be compared with theoretical prediction obtained from Eq.(\ref{eq03}). We present theoretical results from Eq.(\ref{eq03}) in Fig.\ref{fig04}a, where the average capacitance $\alpha$ and its fluctuation are shown as functions of $\eta$ (defined after Eq.(\ref{eq03})). Here $\eta$ plays the role similar to disorder strength. We see that in the clean sample the DOS is large and hence the average capacitance is close to $\langle \alpha \rangle \simeq 1$. In the presence of disorders, the average capacitance decreases with the increasing of $\eta$. The capacitance fluctuation ${\rm rms}(\alpha)$ rises rapidly as the disorder strength is turned on. As $\eta$ is increased further, ${\rm rms}(\alpha)$ reaches a plateau where the magnitude ${\rm rms} (\alpha) \sim 0.2$. When $\eta$ keeps increasing, the fluctuation decays slowly from the plateau value. Our numerical result for $N=1$ is shown in Fig.\ref{fig04}b, with $\langle \alpha \rangle$ and $rms(\alpha)$ changing with disorder strength $W$. Fig.\ref{fig04}a and Fig.\ref{fig04}b show similar behaviors especially the plateau region. From Fig.\ref{fig04}a and Fig.\ref{fig04}b, two differences are noted: (1). the average capacitance in Fig.\ref{fig04}b decays much slower than that in Fig.\ref{fig04}a. (2). the plateau value is about ${\rm rms}(\alpha)=0.32$ at disorder strength $W=7$ which is much larger than that of Fig.\ref{fig04}a. These differences can be understood from the distribution of $\alpha$ shown in Fig.\ref{fig02}. Numerically calculated $P(\alpha)$ has wider spreading in the whole $\alpha$ range than the theoretical function especially at large $\alpha$, featured by the narrow peak at $W=5$ and long tails in $W=9$ and 15. This directly leads to larger average capacitance $\langle \alpha \rangle$ and stronger fluctuation ${\rm rms}(\alpha)$. Our previous analysis suggested that the necklace states are responsible for the peak and long tails in the distribution function at large $\alpha$ area. Here we see that these necklace states greatly enhance the fluctuation of capacitance for the case of single transmission channel ($N=1$).

To make some connection between theoretical result from Eq.(\ref{eq03}) and our numerical calculation, we calculate the mean level spacing $\Delta$ of our system when the lead is removed, which in turn gives the parameter $\eta$ since $\eta=C_e \Delta/e^2$. In Fig.\ref{fig04}c we find that $\eta$ versus disorder strength $W$ showing a nearly linear dependence. From Fig.\ref{fig04}b, we find the maximum value of the plateau locating at $W=7$. The disorder strength $W=7$ corresponds to $\eta=0.57$ from Fig.\ref{fig04}c. Indeed, the maximum plateau value in Fig.\ref{fig04}a is at $\eta=0.57$. Hence our result agrees with the prediction of RMT (Eq.(\ref{eq03})) shown in Fig.\ref{fig04}a, i.e., the capacitance fluctuation reaches its maximum at $W=7$ but with a larger maximum value due to the existance of necklace states.

Since the necklace states enhance the capacitance fluctuation and the probability of occurrence of necklace states is suppressed by increasing the number of conducting channel $N$, we expect that the fluctuation of $\alpha$ at large $N$ should be smaller than the case of $N=1$. We have calculated ${\rm rms}(\alpha)$ for $N=18$ which is shown in Fig.\ref{fig04}d. We see that the fluctuation ${\rm rms}(\alpha)$ vs disorder strength is very similar to that shown in Fig.\ref{fig04}b and the plateau value of ${\rm rms}(\alpha)$ is indeed smaller than that of $N=1$. Interestingly, this plateau value is approaching 0.20, which is theoretically predicted\cite{buttiker96PRL} for $N=1$. Our numerical results show that at even larger $N$, the plateau value of ${\rm rms}(\alpha)$ does not change.

We also noticed that in Fig.\ref{fig04}d the capacitance fluctuation exhibits a plateau structure in disorder interval $W=[15,18]$, with a nearly constant magnitude ${\rm rms}(\alpha) \approx 0.20$ indicating that ${\rm rms}(\alpha)$ is independent of $W$ in this interval. The maximum plateau value is usually identified as the value of universal conductance fluctuation (UCF) in the numerical investigation of UCF in disordered mesoscopic systems.\cite{ren} For instance, the universal spin-Hall conductance fluctuation in the presence of spin orbit interaction was predicted using the maximum plateau value.\cite{ren1} Exact value was later confirmed theoretically by random matrix theory (RMT).\cite{rmt1} Fig.\ref{fig04}d suggests that there may exist a universal behavior of the capacitance fluctuation in the strong disorder regime. Since there is no theoretical prediction of capacitance distribution and its fluctuation for $N \neq 1$, we will carry out extensive numerical calculation for the case of large $N$.

\begin{figure}[tbp]
\includegraphics[width=\columnwidth]{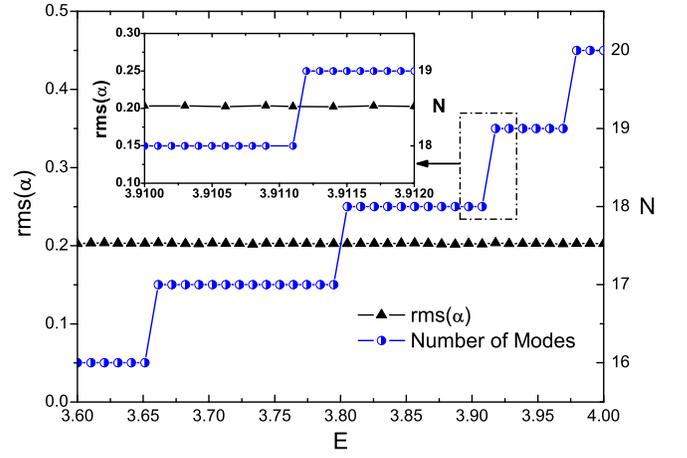}
\caption{The capacitance fluctuation at fixed disorder strength $W=17$
vs Fermi energy. Right axis indicates the channel number in the
system and the inset picture highlights the interval $E\in[3.910,
3.912]$ where $N$ changes from 18 to 19.}\label{fig05}
\end{figure}

Fig.\ref{fig05} plots the maximum value of capacitance fluctuation versus Fermi energy for a fixed disorder strength $W=17$ locating in the middle of the plateau in Fig.\ref{fig04}d, where ${\rm rms}(\alpha)$ is independent of $W$. We have also shown the number of conducting channels for the clean sample vs Fermi energy in the same figure. Clearly the ${\rm rms}(\alpha)$ remains a constant value $0.20$ in this large energy range $E \in [3.6,4]$. This strongly confirms that the fluctuation plateau value saturates at large number of channels. We highlight in the inset of Fig.\ref{fig05} the behavior of ${\rm rms}(\alpha)$ in a small energy interval, corresponding to the number of modes changing suddenly from 18 to 19. Based on these evidences and Fig.\ref{fig04}d, we can safely state that the capacitance fluctuation ${\rm rms}(\alpha)$ is a constant in the disorder window $15 \leq W \leq 18$, and independent of Fermi energy in a wide range $E \in [3.6, 4]$ where $N$ is large. To substantiate the physical picture of universal capacitance fluctuation, we now examine the dependence of capacitance fluctuation on other parameters.

\begin{figure}[tbp]
\includegraphics[width=\columnwidth]{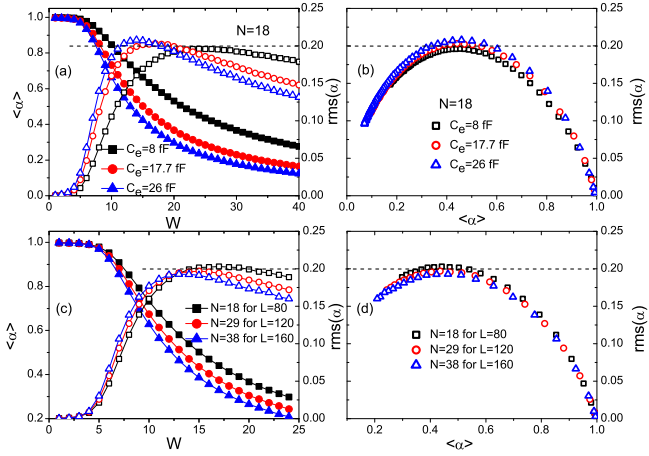}
\caption{The influence of geometric capacitance on electrochemical capacitance fluctuation. Panel (a) shows the $rms(\alpha)$ and $<\alpha>$ vs disorder strength W for different geometric capacitances by varying separation between two plates of the capacitor. Here we have used 80 grid points along each dimension of the mesoscopic capacitor and fixed Fermi energy to be $E=3.85$. Panel (b) is obtained from panel (a) by eliminating $W$ from $rms(\alpha)$ and $<\alpha>$. Panels (c) and (d) are similar to (a) and (b) except that the geometric capacitance is changed by varying the cross-section area of the capacitor. Three systems with width $L$=80,120,and 160 are examined with fixed lattice spacing $a$. Filled symbols represent $<\alpha>$ and unfilled ones correspond to $rms(\alpha)$.}\label{fig06}
\end{figure}

We first check the effect of different geometric capacitances $C_e$ on ${\rm rms}(\alpha)$. In all previous calculations we have chosen $C_e$=17.7 fF for a mesoscopic capacitor with plate size $1600\times1600 nm^2$. Now we investigate the capacitance fluctuation as a function of disorder strength for a series of geometrical capacitances.  Different $C_e$ can be realized by varying the separation between the two plates while keeping the plate area unchanged.\cite{ce} The Fermi energy is fixed at $E=3.85$ ($N=18$ in the lead). The numerical results are depicted in Fig.\ref{fig06}a. We see from Fig.\ref{fig06}a that for different $C_e$ the electrochemical capacitance and its fluctuation are very different as functions of $W$. However, when we plot ${\rm rms}(\alpha)$ versus $<\alpha>$ in Fig.\ref{fig06}b by eliminating the parameter $W$ in Fig.\ref{fig06}a, we find all three curves almost collapse into one with maximum ${\rm rms}(\alpha)\sim 0.20$. This fact suggests that the capacitance distribution does not depend on the geometric capacitance. In Fig.\ref{fig06}c we show the calculated results of electrochemical capacitance by changing the area of the mesoscopic capacitor while fixing the separation between the plates and the lattice spacing $a$. When the area is changed, we keep the ratio of dimension of lead and quantum dot to be $1/4$ so that for a given Fermi energy, larger area contains more subbands in the lead. Three system size with width $L$=80,120, and 160 are examined. In Fig.\ref{fig06}d we plot ${\rm rms}(\alpha)$ versus $<\alpha>$. Here again we see three curves falling into a single one and the maximum capacitance fluctuation close to 0.20. This phenomenon further confirms the conclusion drown from Fig.\ref{fig06}b, the geometric capacitance has no influence on the distribution of $\alpha$. Fig.\ref{fig06}c and Fig.\ref{fig06}d also show that the distribution of electrochemical capacitance and its fluctuation plateau are not affected by the system size.

The behaviors in Fig.\ref{fig06}b and Fig.\ref{fig06}d are very similar to the distribution of conductance in the presence of disorders, where different curves of conductance fluctuation versus average conductance at different energies collapse into a single curve.\cite{qiao10PRB} In Ref.\onlinecite{qiao10PRB} it was found that the universal behavior of conductance fluctuation versus average conductance reflects the universality of conductance distribution: the conductance distribution depends only on the average conductance. Hence our results suggest that the electrochemical capacitance distribution has the following form: $P(\alpha,<\alpha>)$, i.e., the capacitance distribution depends only on the average capacitance. This result is consistent with Eq.(\ref{eq03}), which says that capacitance distribution for $N=1$ depends only on $\beta$ and $\eta$ and so is $<\alpha>=\int d\alpha P(\alpha)$. Eliminating $\eta$ from Eq.(\ref{eq03}) and $<\alpha>$, we find the distribution function as $P(\alpha,\beta,<\alpha>)$.

\begin{figure}[tbp]
\includegraphics[width=\columnwidth]{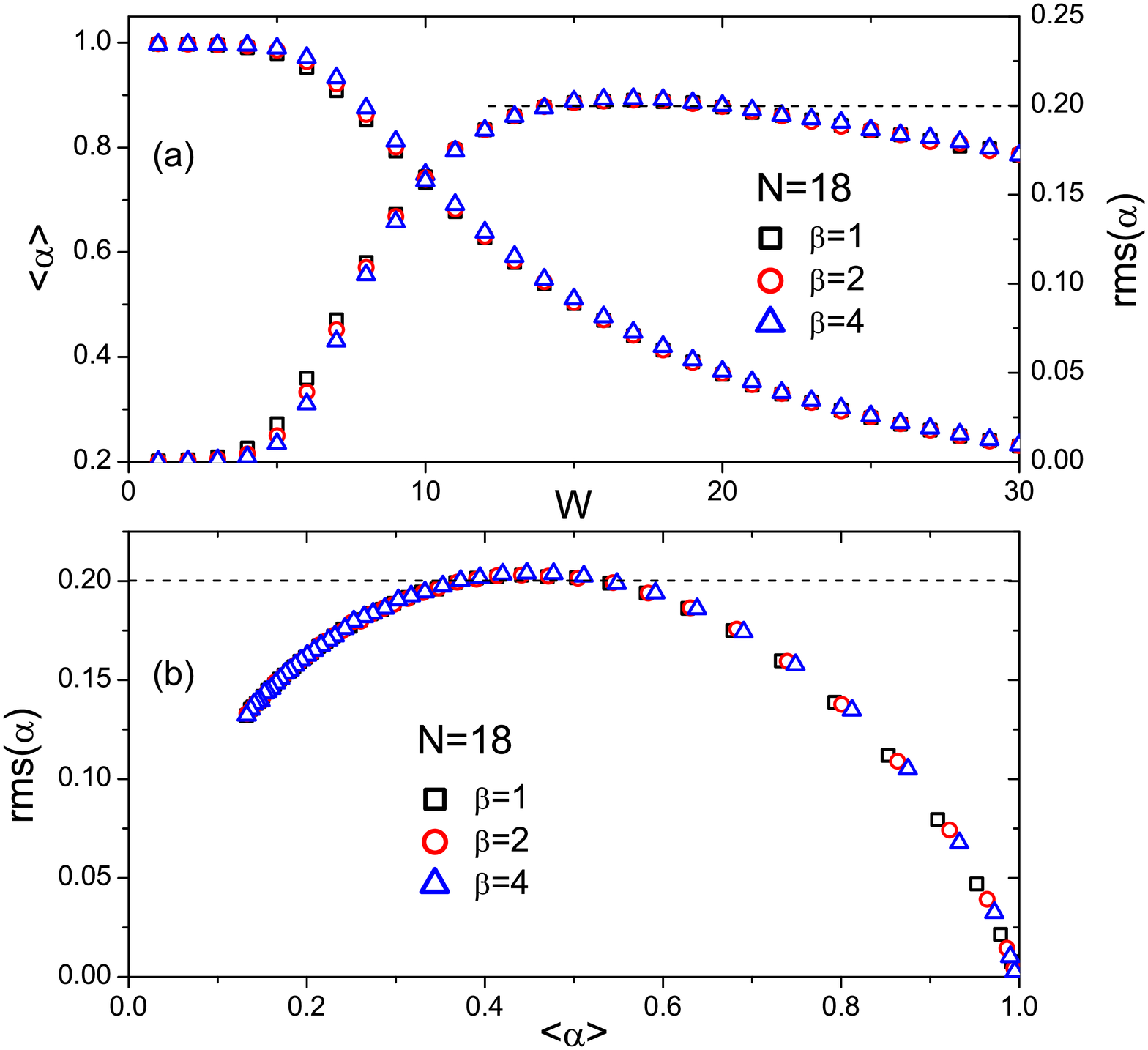}
\caption{Panel (a) shows $rms(\alpha)$ and $<\alpha>$ vs disorder strength W for different ensembles $\beta$ = 1, 2, and 4 at $N=18$. Panel (b) gives the $rms(\alpha)$ versus $<\alpha>$ curves at channel number $N=18$ showing universal behavior for all symmetry classes.}\label{fig07}
\end{figure}

So far, our numerical results show that for the orthogonal ensemble there exists a universal capacitance fluctuation for the electrochemical capacitance in disordered systems, which is independent of disorder strength, Fermi energy, geometric capacitance and system size. This conclusion can be generalized to other symmetry classes. In the previous subsection Sec.\ref{sec31} we already see that the distribution of $\alpha$ in all the three ensembles are exactly the same for large $N$ and $W$. Since the fluctuation ${\rm rms}(\alpha)$ is directly calculated from the distribution function, one expects that the fluctuation of electrochemical capacitance for orthogonal, unitary and symplectic ensembles should be the same. In Fig.\ref{fig07} we provide the numerical evidence to support this point. In Fig.\ref{fig07}a, the average electrochemical capacitance and its fluctuation as a function of disorder strength $W$ for $\beta$ = 1, 2, and 4 are numerically calculated with channel number $N=18$ and other system parameters are kept the same as in Sec.\ref{sec32}. From this figure, we see that three curves are almost identical at large $W$ which is consistent with the observation in Sec.\ref{sec31} and they all develop a plateau behavior with ${\rm rms}(\alpha) \approx 0.20$. For small $W$ between $W=(5,8)$, we do see some difference in three curves. When eliminating $W$ from ${\rm rms}(\alpha)$ and $<\alpha>$, we obtain a perfect single curve in Fig.\ref{fig07}b for all three ensembles. Based on the above analysis, we can conclude that the electrochemical capacitance fluctuation exhibits a universal value $0.20$ at large $N$ for orthogonal, unitary and symplectic classes.

\section{conclusion}\label{sec4}

In conclusion, we perform large scale numerical calculation to investigate the statistical properties of electrochemical capacitance in a disordered mesoscopic capacitor for three different symmetry ensembles. It is found that in the strongly disordered regime, the existence of necklace states greatly influence the statistical behavior of electrochemical capacitance for the case of single conducting channel. Due to the necklace states, the capacitance distribution exhibits a sharply one-sided shape with a nearly-constant tail in large capacitance region at strong disorder, which deviates from theoretical prediction by the random-matrix theory. When the number of conducting channels $N$ is increased, the effect of necklace states gradually diminishes. At large $N$, it is found that in strongly disordered regime the electrochemical capacitance shows universal fluctuation with the magnitude around 0.20, which is independent of the disorder strength, Fermi energy, geometrical capacitance and system size. More importantly, the universal capacitance fluctuation is the same for all three symmetry classes. Finally, the distribution of electrochemical capacitance at large $N$ is found to be a super-universal function that depends only on the average capacitance and does not depend on the symmetry index $\beta$.

\section{acknowledgments}

This work was financially supported by the Research Grant Council (Grant No. HKU 705212P), the University Grant Council (Contract No. AoE/P-04/08) of the Government of HKSAR, and LuXin Energy Group. We thank the information technology services of HKU for providing computational resource on the hpcpower2 system.


\begin{thebibliography}{00}
\bibitem{localization}
It is not straigthforward to calculate the localization length of a quantum dot system. Based on the previous calculation on a 2-dimentional $80 \times 80$ grid, the system enters the localized regime when disorder strength $W$ is larger than 4 for $N$=1 case. The numerical evidence can be found in Ref.[\onlinecite{xu11PRB}]. Hence it is a natural deduction that at larger disorder $W$=15, the quantum dot system is already strongly localized.
\bibitem{ce}
The geometric capacitance in our system can be approximated as that of a parallel plate capacitor, which is $C_e \approx \varepsilon_0 A/d$, where $\varepsilon_0$ is the vacuum permittivity. $A$ and $d$ are the area of the mesoscopic plate and the separation between it and the macroscopic metalic plate, respectively. With fixed area $A=1600\times1600 {nm}^2$, different $C_e$ is acheived by changing the distance $d$. One simply gets that $d=2.84 nm$ corresponds to $C_e=8.0 fF$ and $d=0.87 nm$ for $C_e=26 fF$. Larger capacitance can be realized by sandwiching dielectric material with high permittivity between the two plates.
\end{thebibliography}
\end{document}